\documentclass[fleqn,usenatbib]{mnras}

\usepackage{amssymb}	% Extra maths symbols

\usepackage{newtxtext,newtxmath}
\usepackage[T1]{fontenc}
\usepackage{ae,aecompl}
\usepackage{multirow}
\usepackage{siunitx}
\usepackage[normalem]{ulem}
\usepackage{graphicx}	% Including figure files
\usepackage{amsmath}	% Advanced maths commands

\usepackage[version=4]{mhchem} % chemical formulae
\usepackage{float}	% Positioning of figures - you have added this
\usepackage{stfloats} % Positioning of two column figures - you have added this
\usepackage[table,xcdraw]{xcolor}
\usepackage{booktabs} % for booktabs tables

\title[Variability in observations of Earth-analogue exoplanets]{Variability due to climate and chemistry in observations of oxygenated Earth-analogue exoplanets}

\author[G. J. Cooke et al.]{
G. J. Cooke$^{1}$\thanks{E-mail: pygjc@leeds.ac.uk},
D. R. Marsh$^{1,2}$\thanks{E-mail: d.marsh@leeds.ac.uk},
C. Walsh$^{1}$\thanks{E-mail: c.walsh1@leeds.ac.uk},
S. Rugheimer$^{3,4}$ and G. L. Villanueva$^{5}$
\\
$^{1}$School of Physics and Astronomy, University of Leeds, Leeds, LS2 9JT, UK\\
$^{2}$National Center for Atmospheric Research, Boulder, CO 80301, USA\\
$^{3}$University Oxford, Atmospheric, Oceanic, and Planetary Physics Department, Clarendon Laboratory, Sherrington Road, Oxford OX1 3PU, UK\\ 
$^{4}$Dept of Physics and Astronomy, York University, 4700 Keele Street, North York, Ontario, 3MJ 1P3, Canada\\ 
$^{5}$NASA Goddard Space Flight Center, Solar System Exploration Division, 8800 Greenbelt Road, Greenbelt, MD 20771, USA\\
}

\date{Accepted XXX. Received YYY; in original form ZZZ}

\pubyear{2022}

\begin{document}
\label{firstpage}
\pagerange{\pageref{firstpage}--\pageref{lastpage}}
\maketitle

% Abstract of the paper
\begin{abstract}

\noindent The Great Oxidation Event was a period during which Earth's atmospheric oxygen (\ce{O2}) concentrations increased from $\sim 10^{-5}$ times its present atmospheric level (PAL) to near modern levels, marking the start of the Proterozoic geological eon 2.4 billion years ago. Using WACCM6, an Earth System Model, we simulate the atmosphere of Earth-analogue exoplanets with \ce{O2} mixing ratios between 0.1\% and 150\% PAL. Using these simulations, we calculate the reflection/emission spectra over multiple orbits using the Planetary Spectrum Generator. We highlight how observer angle, albedo, chemistry, and clouds affect the simulated observations. We show that inter-annual climate variations, as well short-term variations due to clouds, can be observed in our simulated atmospheres with a telescope concept such as LUVOIR or HabEx. Annual variability and seasonal variability can change the planet's reflected flux (including the reflected flux of key spectral features such as \ce{O2} and \ce{H2O}) by up to factors of 5 and 20, respectively, for the same orbital phase. This variability is best observed with a high-throughput coronagraph. For example,  HabEx (4 m) with a starshade performs up to a factor of two times better than a LUVOIR B (6 m) style telescope. The variability and signal-to-noise ratio of some spectral features depends non-linearly on atmospheric \ce{O2} concentration. This is caused by temperature and chemical column depth variations, as well as generally increased liquid and ice cloud content for atmospheres with \ce{O2} concentrations of $<$1\% PAL.

\end{abstract}

% Select between one and six entries from the list of approved keywords.
% Don't make up new ones.

%%%%%%%%%%%%%%%%%%%%%%%%%%%%%%%%%%%%%%%%%%%%%%%%%%

%%%%%%%%%%%%%%%%% BODY OF PAPER %%%%%%%%%%%%%%%%%%

\section{Introduction}
\label{Introduction section}

During the current era of exoplanet discovery, the search for a temperate Earth-sized exoplanet orbiting around a G-type star continues. Data from telescopes and future observatories will add thousands more exoplanets to the list of over 5000 that have been discovered to date\footnote{\href{https://exoplanetarchive.ipac.caltech.edu/}{https://exoplanetarchive.ipac.caltech.edu/}}$^{,}$\footnote{\href{http://www.exoplanet.eu/}{http://www.exoplanet.eu/}}. Upcoming transit and radial velocity surveys will make more precise measurements for the masses and radii of terrestrial exoplanets, thus enabling confident estimates of their bulk density and surface gravity. In the future, an Earth-sized exoplanet orbiting a G-type star at a semi-major axis of approximately 1 au may be found. This poses the question: will it be habitable? 

Atmospheric information can be gained from measurements of emission, reflection or transmission spectroscopy. Exoplanet atmospheres have been observed and characterised since the early 2000's \citep{Charbonneau_2002}, with larger, Jupiter-sized planets being the least challenging to observe \citep{2010ARA&A..48..631S}. Detecting atmospheric properties and molecules for an Earth-sized exoplanet around a G-type star is much more challenging due to the comparatively weaker spectroscopic signal.

Any measurement of an exoplanet's atmosphere will be taken during a specific period of its geological evolution \citep{2007ApJ...658..598K}. The Earth itself has had a complicated geological and biological history which is still not wholly understood \citep{2020SciA....6.1420C, 2020PreR..343j5722S,cole2020co}. At least three geological eons have harboured life on Earth: the Archean (4-2.4 Gyr ago), the Proterozoic (2.4-0.54 Gyr ago) and the Phanerozoic \citep[0.54 Gyr ago - present;][]{2009Natur.461..179L, betts2018integrated, dodd2017evidence}. The atmosphere during the Archean was anoxic, likely containing larger amounts of \ce{CO2} and \ce{CH4} compared to present-day \citep[see Table~\ref{Chemical table};][]{2008AsBio...8.1127H, 2020SSRv..216...90C}. The emergence of cyanobacteria, in addition to hydrogen escape throughout the Archean \citep{2001Sci...293..839C,2019GeCoA.244...56Z}, have been proposed to have led to the Great Oxidation Event (GOE) approximately 2.4 Gyr ago \citep{2020PNAS..11713314W}. The GOE enabled oxygen to increase to concentrations $\geq$0.1\% the present atmospheric level \citep[PAL;][]{2002Sci...296.1066K}, with possible Proterozoic eon concentrations lying between $1$ PAL and $10^{-3}$ PAL \citep{2020PreR..343j5722S,2021AsBio..21..906L}. Phanerozoic \ce{O2} concentrations are posited to have been between $\sim$0.1 PAL and $\sim$2 PAL \citep{2018haex.bookE.189O, 2019MinDe..54..485L}. Table~\ref{Chemical table} presents a simplified overview of the composition of Earth's atmosphere for the past 4 billion years. These three geological eons represent important observational windows for Earth-analogue exoplanets \citep{2018ApJ...854...19R} and the search for habitable and inhabited exoplanets.

\begin{table}
\caption{A simplified chemical timeline of Earth is presented for \ce{O2}, \ce{CO2}, and \ce{CH4}, during the Archean, mid-Proterozoic and Phanerozoic. Molecular abundances are constraints from years of research, including geological evidence and modelling studies. The `inclusive' data from \protect\cite{2018haex.bookE.189O} is listed. Data for \ce{O2} and \ce{CH4} during the mid-Proterozoic is also from \protect\cite{2019MinDe..54..485L}, \protect\cite{2020PreR..343j5722S}, and \protect\cite{2021AsBio..21..906L}. The mixing ratio for \ce{O2} is given relative to its present atmospheric level (PAL) of 0.21. The mixing ratios for \ce{CO2} and \ce{CH4} are given in terms of parts per million by volume (ppmv). 1 Ga represents 1 billion years in the past.}
\label{Chemical table}
\begin{center}
\begin{tabular}{@{}ccccc@{}}
\toprule \midrule
\begin{tabular}[c]{@{}c@{}}Molecular\\abundance\end{tabular} & \begin{tabular}[c]{@{}c@{}}Archean\\4 - 2.4 Ga\end{tabular} & \begin{tabular}[c]{@{}c@{}}mid-Proterozoic\\1.8 - 0.8 Ga\end{tabular} & \begin{tabular}[c]{@{}c@{}}Phanerozoic\\0.54 - 0 Ga\end{tabular} \\ \midrule
\ce{O2} [PAL]  & $10^{-12} - 10^{-5}$ & $10^{-5} - 1$  & $10^{-1} - 2$ \\ %
\ce{CO2} [ppmv] & $2500-40,000$ & $1400-28,000$  & $200-2,800$ \\
\ce{CH4} [ppmv] & $100-35,000$ & $1-100$ & $0.4-10$\\ 
\bottomrule
\end{tabular}
\end{center}
\end{table}

Gases such as \ce{O2}, \ce{N2O}, and \ce{CH4}, are primarily produced by life on Earth. However, these gases can also be abiotically produced \citep{2014ApJ...785L..20W,2018ApJ...862...92K} and the detection of such gases in an exoplanet atmosphere is not a conclusive detection of the presence of life \citep{2002A&A...388..985S, 2017PhR...713....1G}. \ce{O3} is the photochemical byproduct of \ce{O2} and therefore indicates the presence of \ce{O2} without a detection of \ce{O2} itself. Several modeling studies have shown that the amount of \ce{O3} in an Earth-analogue atmosphere increases as the concentration of \ce{O2} rises \citep{1980JGR....85.3255K,2003AsBio...3..689S,2017ApJS..231...12W, 2021E&PSL.56116818G, 2022RSOS....911165C}. \cite{2018ApJ...858L..14O} discussed how seasonal \ce{O2} variations could give rise to seasonal changes in the reflectance spectra of \ce{O3} in the UV region, noting however that separate hemispheres that experience opposite seasons mask the full seasonality signature. 

Previous work has modelled modern Earth as an exoplanet \citep{2009ApJ...698..519K}, as well as Earth at different geological time periods \citep{2015ApJ...806..137R, 2018ApJ...854...19R, 2016AsBio..16..873A, 2020ApJ...892L..17K, 2020ApJ...904...10K}. Other studies investigated the photometric and spectroscopic variations of an Earth-analogue exoplanet \citep{2001Natur.412..885F, 2006ApJ...651..544M,2006AsBio...6...34T,2006AsBio...6..881T}. These investigations have used a variety of 1D and 3D models, and it is clear that numerous parameters, including geological eras with different atmospheric compositions, influence molecular detectability. The emission and transmission spectra of oxygenated exoplanets (varying through geological time) has been predicted with 1D models \citep[e.g. ][]{2003AsBio...3..689S,2018ApJ...854...19R,2020ApJ...892L..17K}. Whilst most work in this area has focused on modelling, \cite{2022AJ....163....5D} performed a phase angle analysis of Earth using data from the EPOXI mission which observed Earth to test exoplanet observational predictions \citep{2011AsBio..11..907L}. 

Emission and reflectance spectra can be observed through high-contrast imaging (HCI). This does not require the planet to transit the star, although the observer's inclination to the plane of the orbit will affect the observations because the brightness amplitude is reduced as the orbital inclination decreases from "edge-on" to "face-on" \citep{2012A&A...538A..95M}. As the planet orbits, it reflects light, particularly in the ultraviolet and visible wavelength regions. The Earth's emission peaks in the infrared, where the emission (both in terms of its strength and wavelength variance) depends on temperature, clouds, and chemistry \citep{2018haex.bookE.116P}. The brightness of the host star (G-type) far outshines the brightness of an Earth-analogue exoplanet. Coronagraphs can be used to reduce the incoming light from the star, with a contrast of $\sim 10^{-10}$ required to reliably characterise an Earth-Sun analogue system \citep{2018Galax...6...51L, 2022arXiv220505696C}; hence, we explore the use of a coronograph in this work. 

Recently, \cite{2021AJ....161..150C} evaluated the detectability of \ce{O3} and \ce{O2} on Earth-analogue (defined as possessing an oxygenated atmosphere due to life) exoplanets with four different telescope concepts: LUVOIR A (15 m diameter), LUVOIR B (8 m diameter), HabEx (4 m diameter) with and without a starshade. They simulated observations with these missions for \ce{O2} concentrations of $10^{-5}$ PAL and upwards to 1 PAL, finding that larger diameter telescopes will better constrain the frequency of Earth-analogue worlds, noting that this depends on the occurrence rate of exo-Earth candidates.

In this work, we present thousands of synthetic high-contrast imaging spectra based on simulations of Earth-analogue exoplanet atmospheres over various atmospheric oxygen concentrations. This enables us to predict the variability in future observations, both within spectral features of interest, and in the broadband. These atmospheric simulations utilise a 3D Earth System Model with fully coupled chemistry and physics - the Whole Atmosphere Community Climate Model 6 (WACCM6). Using the Planetary Spectrum Generator (PSG), we investigate the impact that changing oxygen concentrations has on simulated observations, considering climate, chemistry, temporal, annual, and seasonal variations. We determine the observability of such variations with next generation telescope concepts.

\section{Methods}
\label{Methods section}

\subsection{WACCM6}
\label{The climate model WACCM6 section}

WACCM6 is an Earth System Model configuration of the Community Earth System Model (CESM) which couples together atmosphere, land, land-ice, ocean and sea-ice sub models \citep{2019JGRD..12412380G}. The model release used in this work was the Coupled Earth System Model version 2.1.3 (CESM2.1.3\footnote{\href{http://www.cesm.ucar.edu/models/cesm2/}{http://www.cesm.ucar.edu/models/cesm2/}}). The configuration that we used (BWma1850 compset\footnote{A list of compsets can be found here: 

\href{https://www.cesm.ucar.edu/models/cesm2/config/2.1.3/compsets.html}{https://www.cesm.ucar.edu/models/cesm2/config/2.1.3/compsets.html}}) has 98 chemical species, 208 chemical reactions, and 90 photolysis reactions.  The model resolution was $1.875^{\circ}$ in latitude by $2.5^{\circ}$ in longitude.

We simulated five different scenarios that are presented in Table~\ref{run table}. The scenarios were: a pre-industrial atmosphere simulation (hereafter PI), and simulations with atmospheres that have \ce{O2} concentrations of 150\% PAL, 10\% PAL, 1\% PAL, and 0.1\% PAL. Each simulation had a 24 hr rotation rate and the modern Earth's land and ocean configuration. Each simulation was stopped when atmospheric trends in temperature and chemistry were determined to have fluctuations that are expected from a climate model (e.g. year-to-year and diurnal cycle variations) for the last four model years, rather than ongoing trends (e.g. a decrease in surface temperature year-on-year). 

The mixing ratios of the following species were held constant at the surface: \ce{O2} (varied as in Table \ref{run table}), \ce{CH4} (0.8 ppmv), \ce{CO2} (280 ppmv), \ce{N2O} (270 ppbv), and \ce{H2} (500 ppbv). The fixed mixing ratio of other minor constituents (mixing ratio $<1$ ppbv) are listed on the Exo-CESM GitHub\footnote{\href{https://github.com/exo-cesm/CESM2.1.3/tree/main/O2_Earth_analogues}{https://github.com/exo-cesm/CESM2.1.3/tree/main/O2\textunderscore Earth\textunderscore analogues}} page. The chemistry of all other species was left to evolve according to the chemical mechanism. All simulated atmospheres have a surface pressure of 1,000 hPa. In each simulation in which the \ce{O2} concentration was decreased, the \ce{N2} concentration was increased to maintain a 1,000 hPa surface pressure. Each simulation used the modern-day Sun as the solar input spectrum. This solar spectrum's irradiance was specified between 10.5 nm and 99975.0 nm.

Model data was output as mean average values (the sum of the instantaneous output at each time step divided by the number of time steps considered) every 5 days, and mean average values every month. For the mean vertical profiles computed for Fig.~\ref{Chemistry figure}, time and longitudinal means were taken, and latitudinal weights were averaged over (grid boxes near the poles have a smaller area than grid boxes near the equator), leaving just the vertical level dimension. Additionally, the instantaneous model state was output every 5 days (a "snapshot" of the model). The snapshots were used for estimating reflection/emission spectra at a specific point in time. Time means presented in Fig.~\ref{Chemistry figure} and Table~\ref{run table} were averaged over the last 4 years of each simulation. 

\subsection{Planetary Spectrum Generator}
\label{PSG section}

\begin{table*}
\caption{Annual mean modelled temperatures and ozone columns under five different \ce{O2} mixing ratio ($f_{\textrm{O}_2}$)  scenarios. $f_{\textrm{O}_2}$ is given in terms of present atmospheric level (PAL = 21\% by volume of the atmosphere). Listed are the global mean surface temperature, $\overline{T}_\textrm{S}$, the global mean tropopause temperature ($\overline T_\textrm{Tp}$), and the minimum ($T_\textrm{Tp,min}$) and maximum values ($T_\textrm{Tp,max}$) of the global tropopause temperature. Also listed is the global mean ozone column, $\overline C_{\textrm{O}_3}$, as well as the maximum $C_{\textrm{O}_3, \textrm{max}}$, and minimum, $C_{\textrm{O}_3, \textrm{min}}$, ozone columns in Dobson Units (DU), where 1 DU = $2.687\times10^{20}$ $\textrm{molecules} \cdot \textrm{m}^{-2}$. All temperatures and \ce{O3} columns  are time-averaged over the final 4 years of each simulation.} 
\label{run table}
\begin{center}
\begin{tabular}{@{}cccccccccc@{}}
\toprule \midrule
Simulation name & $f_{\textrm{O}_2}$ [PAL] & $\overline{T}_\textrm{S}$ [K] & $\overline{T}_\textrm{Tp}$ [K] & $T_\textrm{Tp, min}$ [K] &$ T_\textrm{Tp, max}$ [K]& $\overline C_{\textrm{O}_3}$ [DU] & $C_{\textrm{O}_3, \textrm{min}}$ [DU] & $C_{\textrm{O}_3, \textrm{max}}$ \\ \midrule
     150\% PAL & 1.500 & 288.5 & 205.0 & 193.6 & 216.3 & 313 & 261 & 426 \\   
     PI        & 1.000 & 288.5 & 204.4 & 193.3 & 216.0 & 279 & 236 & 375 \\  
     10\% PAL  & 0.100 & 288.9 & 204.9 & 194.0 & 215.7 & 169 & 130 & 235 \\  
     1\% PAL   & 0.010 & 288.5 & 202.4 & 191.6 & 213.5 & 66  & 41  & 83  \\  
     0.1\% PAL & 0.001 & 286.0 & 196.6 & 184.4 & 210.0 & 18  & 13  & 23  \\  
\bottomrule
\end{tabular}
\end{center}
\end{table*}

The Planetary Spectrum Generator \citep[PSG;][]{2018JQSRT.217...86V} was used to compute theoretical relection/emission spectra for the WACCM6 atmospheric simulations. We used the GlobES 3D mapping tool to upload 3D data from our simulated atmospheres\footnote{ \href{https://psg.gsfc.nasa.gov/apps/globes.php}{https://psg.gsfc.nasa.gov/apps/globes.php}}. To reduce the memory footprint of the model input to be compatible with GlobES in PSG, the data were rebinned to a resolution of $10^{\circ}$ in longitude only (with the latitudinal grid resolution unchanged at a value of $1.875^{\circ}$).

All of the synthetic spectra were calculated assuming an Earth-analogue exoplanet around a G2V star with a 1-year orbit at a $90^\circ$ inclination angle. Edge on systems are also likely to be best characterised in terms of mass and radius. All synthetic spectra were set to an exposure time of 60 seconds, with 1440 exposures, adding up to a total integration time of 24 hr. Although the noise estimates do not simply scale with distance because of reduced coronagraphic throughput at larger distances, to good approximation, the noise estimates do scale with total integration time, $T_\textrm{int}$ ($\sigma_\textrm{noise} \propto T_\textrm{int}^{0.5}$). We show spectra for an integration time of 48 hr, where we assume that the noise is reduced by a factor of $\sqrt{2}$ from the 24 hr PSG results. Synthetic spectra from HCI observations were computed with WACCM6 snapshots every 5 days at 00:00 UTC, from January 1$^\textrm{st}$ until December 27$^\textrm{th}$. Consequently, the illuminated side of the planet is centred at $180^\circ$ longitude, where it is predominantly ocean (i.e. it is mostly the Pacific ocean that is receiving sunlight at these times). Each snapshot was ingested into PSG by updating the orbital parameters for Earth (using the ephemeris data from the year 2020) every 5 days. This was then matched with the longitude and latitude of the Earth grid, to simulate the orbit in PSG to coincide with the orbit of the WACCM6 simulations.

The molecules used for the computation of reflection/emission spectra were \ce{N2}, \ce{O2}, \ce{CO2}, \ce{H2O}, \ce{O3}, \ce{CH4}, and \ce{N2O}. Each molecule used the default HITRAN opacity data \citep{2017JQSRT.203....3G}. All available collision induced absorption coefficients were utilised. The effects of scattering were included, as well as that of ice clouds and water clouds \citep{2011Icar..216..227V}. The effective radius of the cloud particles was assumed to be \SI{5}{\micro \metre} for water (stratocumulus) clouds and \SI{100}{\micro \metre} for ice (cirrus) clouds, as is typical in the modern Earth's atmosphere. We also ran the same set of simulations with smaller diameter cloud particles, assuming \SI{1}{\micro \metre} for water clouds and \SI{1}{\micro \metre} for ice clouds - equivalent figures to figures 3-7 are shown in the Supplementary Information.

Telescope templates are used for the Large Ultraviolet Optical Infrared Surveyor \citep[LUVOIR;][]{2017SPIE10398E..09B} concept and the Habitable Exoplanet Observatory (HabEx) concept. For LUVOIR, we used both the A (15 m primary mirror) and B (8 m primary mirror) templates. We changed the B template to a 6 m primary mirror based on the recent suggestions from the Decadal Survey \citep{2021pdaa.book.....N}. The LUVOIR A and B concepts have spectral resolutions of $R=7$ (0.2-\SI{0.515}{\micro \metre}), $R=140$ (0.515-\SI{1.0}{\micro \metre}) and $R=70$ (1.01-\SI{2.0}{\micro \metre}) for the UV, VIS and NIR channels respectively (where $R=\lambda / \Delta \lambda$, $\lambda$ is the wavelength, and $\Delta \lambda$ is the wavelength bin width). The HabEx concept had the same spectral resolutions in the UV (0.2-\SI{0.45}{\micro \metre} and VIS channels (0.45-\SI{0.975}{\micro \metre}) but instead with $R=40$ in the NIR channel (0.975-\SI{1.8}{\micro \metre}).

Idealised spectra were computed with PSG (Fig.~\ref{Ideal emission spectra figure}). However, to deduce whether atmospheric constituents in any of the WACCM6 simulations are physically observable, we also estimated the atmospheric signal-to-noise ratio, $\textrm{SNR}$, on the produced spectra. PSG calculated the noise, $\sigma_\textrm{noise}$, on an observation set. $\sigma_\textrm{noise}$ included contributions from the source (the star and planet), from the background sky, and from the telescope itself (read noise and dark noise).

The SNR of a molecular detection also depends on a `baseline' atmosphere which does not contain that molecule. This provides a continuum with which to estimate the absorption strength of the spectral features. We use a 1,000 hPa \ce{N2} atmosphere as our baseline which was created from each simulation by removing all other atmospheric molecules, but keeping the presence of liquid and ice clouds in the atmosphere. SNR were calculated at a specific wavelength with no rebinning performed on the simulated spectra.

The Python code to produce configuration files to upload to PSG is included with the data associated with this manuscript. We note here that all simulations with PSG computed for this work are forward models. More details on the simulations completed are given in Appendix \ref{Appendix PSG simulations section}.

\section{Results}

\subsection{WACCM6 simulations}
\label{Atmospheric results from the simulations section}

Here we summarise the globally-averaged and time-averaged atmospheric chemistry of the simulations that were fed into PSG. See \cite{2022RSOS....911165C} for a more detailed discussion.

\begin{figure*}
	\centering
	\includegraphics[width=\textwidth]{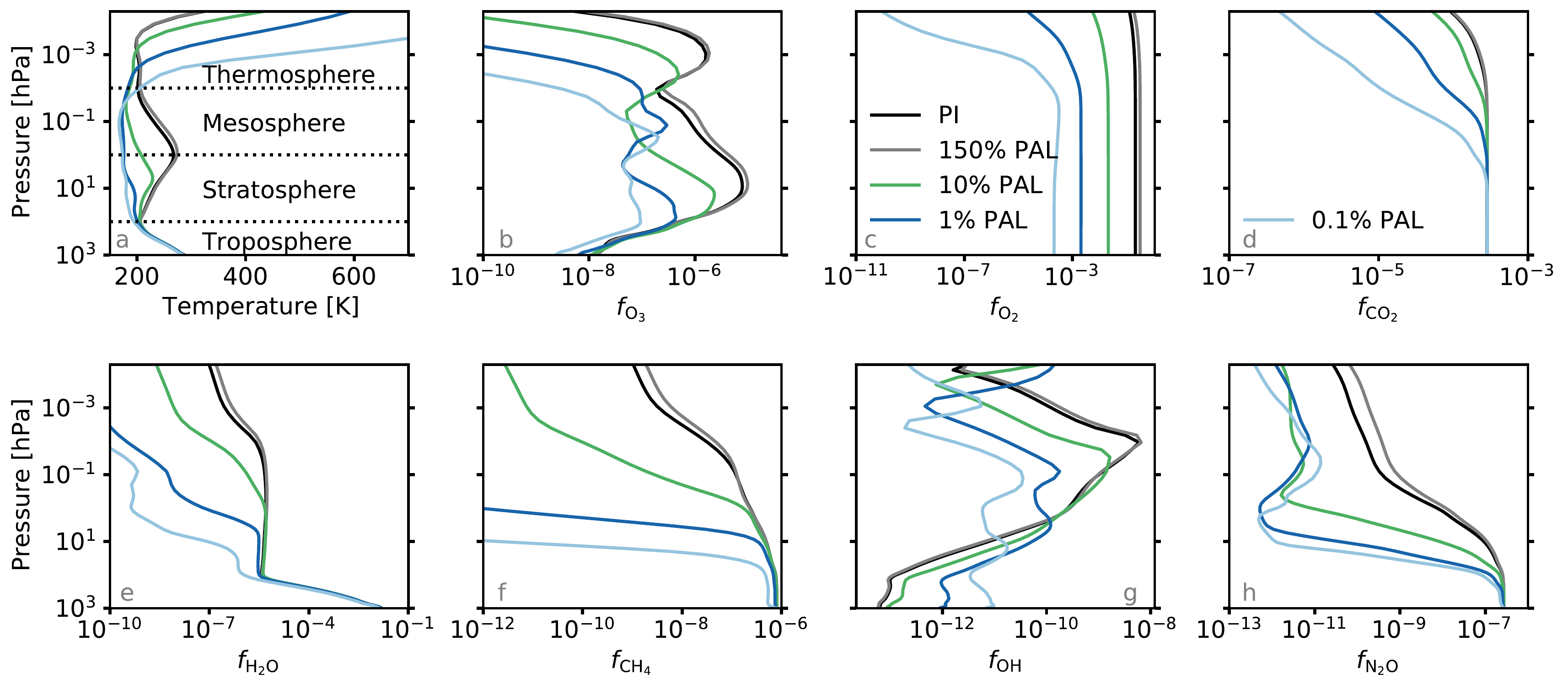}
    \caption{(\textbf{a}) The vertical profile of global mean temperatures for the following simulations: 150\% PAL (grey), PI (black), 10\% PAL (green), 1\% PAL (dark blue), and 0.1\% PAL (light blue) \ce{O2} atmospheres. Black dashed lines indicate the transitions between the major temperature layers in the pre-industrial atmosphere. Mixing ratio vertical profiles, denoted as $f_X$, where $X$ is the molecule shown, are plotted for \ce{O3} (\textbf{b}), \ce{O2} (\textbf{c}), \ce{CO2} (\textbf{d}), \ce{H2O} (\textbf{e}), \ce{CH4} (\textbf{f}), \ce{OH} (\textbf{g}) and \ce{N2O} (\textbf{h}).}
    \label{Chemistry figure}
\end{figure*}

Fig.~\ref{Chemistry figure} shows the atmospheric mixing ratio vertical profiles from all five WACCM6 simulations for \ce{O3, O2, CO2, H2O, CH4, OH} and \ce{N2O}. Also shown are the vertical profiles for temperature. When \ce{O2} is reduced, there is less \ce{O} produced from the photodissociation of \ce{O2}, such that fewer \ce{O3} molecules can be produced from the reaction 
\begin{equation}
    \ce{O + O2 + M -> O3 + M},
    \label{equation 2}
\end{equation}
\noindent where M is any third body. 

When the \ce{O2} concentration decreases (from 150\% PAL down to 0.1\% PAL), the \ce{O3} column also decreases. Whilst this effect has been reported in previous work, our WACCM6 simulations result in \ce{O3} columns that are reduced by up to a factor of $\sim5$ compared to previous work for \ce{O2} concentrations between 0.5\% and 50\% PAL \citep[see][]{2022RSOS....911165C}. With less \ce{O3} to absorb ultraviolet (UV) radiation, at $\leq1\%$ PAL, there is reduced heating in the lower stratosphere and this cools the tropopause. Hence, more water is frozen out in the tropopause `cold-trap' \citep{2009RvGeo..47.1004F}, generally resulting in more liquid and ice clouds.

Increased photolysis rates for \ce{CH4, N2O} and \ce{CO2} result in lower mixing ratios. The mixing ratios for \ce{N2O} and \ce{CO2} are reduced by a factor of 100 or more by the mesopause, and that for \ce{CH4} is reduced by several orders of magnitude above the tropopause. Furthermore, as \ce{O2} reduces, \ce{H2O} is increasingly photolysed above the tropopause and has mixing ratios reduced below 0.1 ppmv (parts per million by volume) in the stratosphere for the 1\% PAL and 0.1\% PAL cases, or in the upper mesosphere for the 10\% PAL case. This is in contrast to the PI atmosphere that retains at least 0.1 ppmv of \ce{H2O} in the lower thermosphere. Therefore, the five major greenhouse gases in Earth's atmosphere are all reduced above the tropopause when molecular oxygen is reduced. In addition to reduced heating from UV absorption by \ce{O3}, this results in cooling of the troposphere in the simulations with the lowest \ce{O2} concentrations, resulting in a reduced water vapour column. 

% Chemistry figure

% Ideal emission spectra figure
\begin{figure}
	\centering
	\includegraphics[width=\columnwidth]{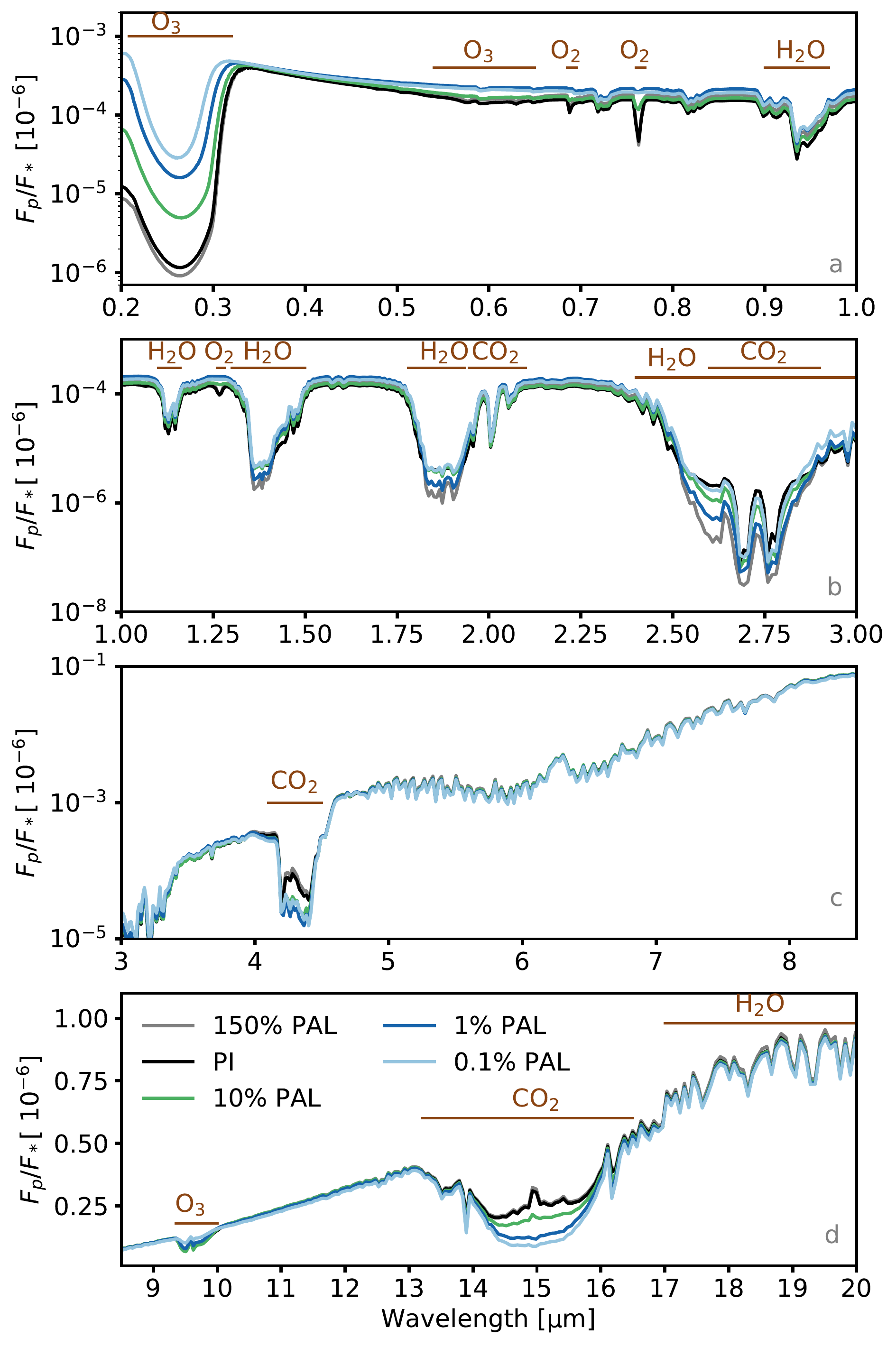}
    \caption{Idealised reflection/emission spectra in which a coronagraph is not used, given in terms of planet-to-star flux ratio ($F_p/F_*$ [$10^{-6}$]), for the 150\% PAL (grey), PI (black), 10\% PAL (green), 1\% PAL (dark blue), and 0.1\% PAL (light blue) atmospheres between \SI{0.1}{\micro \metre}-\SI{1}{\micro \metre} (\textbf{a}), \SI{0.1}{\micro \metre}-\SI{3}{\micro \metre} (\textbf{b}), \SI{3}{\micro \metre}-\SI{8.5}{\micro \metre} (\textbf{c}), and \SI{8.5}{\micro \metre}-\SI{20}{\micro \metre} (\textbf{d}). Key spectral features are indicated. The spectra are computed at a phase of $395.5^\circ$ for climate simulations at midnight on March 22$^\textrm{nd}$. Note that each subplot has a different $y$ axis range and scaling. The spectra are binned to a spectral resolution of $R=250$, where $R=\lambda/\Delta\lambda$, $\lambda$ is the wavelength of light, and $\Delta\lambda$ is the wavelength bin width.}
    \label{Ideal emission spectra figure}
\end{figure}

\subsection{PSG reflection and emission spectra output}

Fig.~\ref{Ideal emission spectra figure} shows reflection and emission spectra for the PI, 150\% PAL, 10\% PAL, 1\% PAL, and 0.1\% PAL atmospheres. These spectra are simulated on the 22$^\textrm{nd}$ of March, which is at $359.5^\circ$ phase in our geometrical setup, just before the exoplanet goes into secondary eclipse at $0^\circ$ phase. In reality, this idealised spectrum will not be observable because of the small angular separation at this phase that results in low coronagraph throughput. 

The emission from several spectral features are modified when oxygen is reduced. For example, the peak at \SI{15}{\micro\metre} for \ce{CO2}, within an absorption trough, indicates the presence of a hot stratosphere, as has previously been reported \citep{2000ESASP.451..133S,2007ApJ...658..598K,2018ApJ...854...19R}. Once \ce{O2} has reduced to $\leq1$\% PAL in our simulations, the stratosphere no longer exists, so this feature is seen completely in absorption. \ce{O2} features seen in absorption are at \SI{0.69}{\micro\metre}, \SI{0.76}{\micro\metre}, and \SI{1.27}{\micro\metre}. An increase in oxygen increases the depth of these features relative to the continuum. There is a noticeable difference in magnitude between the PI and 150\% PAL cases; the 150\% PAL \ce{O2} features are 13\%, 9\% and 28\% deeper relative to the continuum at $R=250$, respectively. When \ce{O2} is reduced, there is less stratospheric \ce{O3} resulting in a lower total \ce{O3} column; hence, there is reduced atmospheric absorption by \ce{O3}. This is most noticeable at UV wavelengths, where the depth of the feature lies between $\approx9\times10^{-7}$ ppm to $\approx3\times10^{-5}$ ppm. The depth of \ce{H2O} features mostly result from the climate at the chosen time stamp and orbital phase, although the time-averaged water column is reduced by up to 20\% at 0.1\% PAL because of lower tropospheric temperatures.

The spectra in Fig.~\ref{Ideal emission spectra figure} are simulated at a single point in time. However, the spectra will change throughout the orbit, due to seasonal, climatic, and chemical changes in the atmosphere. The presence and distribution of clouds also affects reflection and emission spectra. Clouds can either mask molecular features, or boost them through reflectivity \citep{2018ApJ...854...19R}. We now simulate the spectra of these exoplanets as seen through HCI by sampling the climate system along the orbit. We then use these simulations to estimate the variations expected from future observations of oxygenated Earth-analogue exoplanets.

\subsection{Spectra from high-contrast imaging}

% SNR bar chart figure
\begin{figure*}
	\centering
	\includegraphics[width=1\textwidth]{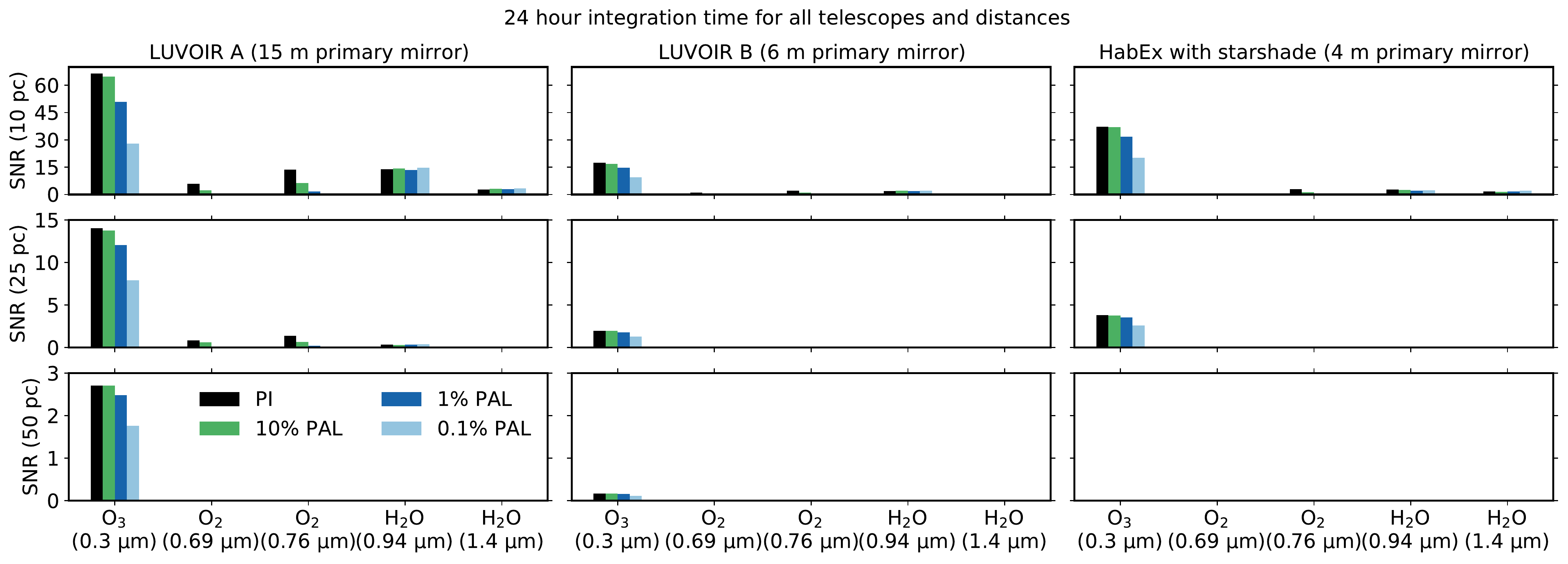}
    \caption{The signal-to-noise ratio for five molecular features are given for the PI (black), 10\% PAL (green), 1\% PAL (dark blue), and 0.1\% PAL (light blue) atmospheres, at a distance of 10 pc (\textbf{top row}), 25 pc (\textbf{middle row}), and 50 pc (\textbf{bottom row}). The LUVOIR A (\textbf{left column}), LUVOIR B (\textbf{middle column}), and HabEx with a starshade (\textbf{right column}) telescope concepts are evaluated, each observing for a total integration time of 24 hr. Only the first year of the final four years of each simulation is evaluated in this figure.}
    \label{SNR bar chart figure}
\end{figure*}

% Annual variability figure
\begin{figure*}
	\centering
	\includegraphics[width=1\textwidth]{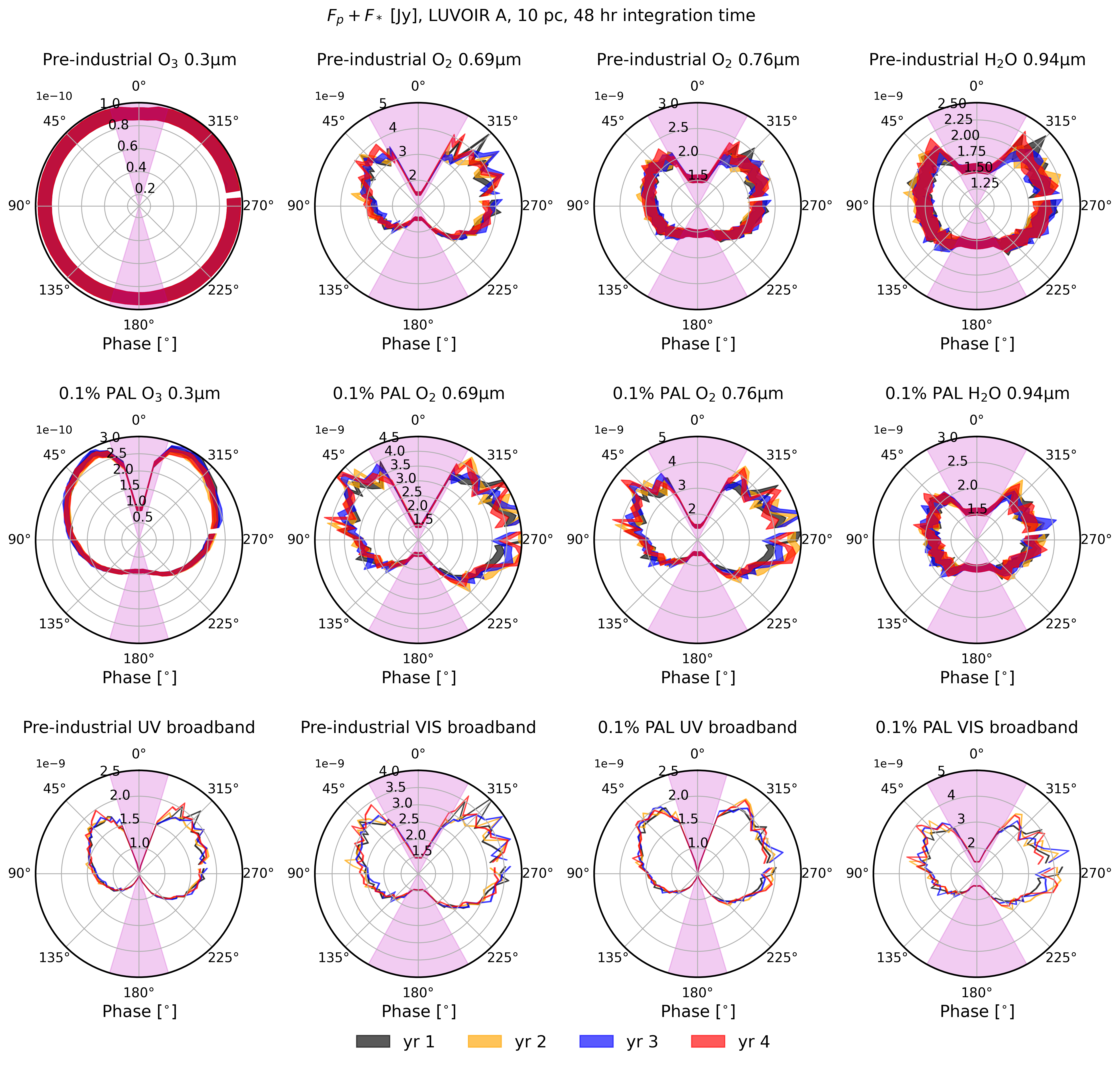}
    \caption{The total flux measured as a function of orbital phase for four consecutive years is plotted to demonstrate year-on-year variability.  Plotted in the radial direction in units of Jy is the total flux $F_T = F_p + F_*$ from the planet and star as seen by the detector from the final 4 years (1st year = black, 2nd year = yellow, 3rd year = blue, 4th year = red) of the PI and 0.1\% PAL simulated atmospheres. Note that the total flux never goes to zero because there is always some flux from the star. The total integration time is 48 hr using LUVOIR A for an exoplanet at 10 pc. The discontinuity in the curves is between December $27^\textrm{th}$ and January $1^\textrm{st}$. \textbf{(Top and Middle)} The top and middle rows show the variation with phase (in the azimuthal direction) at the \SI{0.3}{\micro \metre} \ce{O3} feature, at the \SI{0.69}{\micro \metre} \ce{O2} feature, at the \SI{0.76}{\micro \metre} \ce{O2} feature, and at the the \SI{0.94}{\micro \metre} \ce{H2O} feature for the PI and 0.1\% PAL atmospheres. The width of the lines indicate the noise for that particular measurement. \textbf{(Bottom)} The bottom row shows the broadband variation in the PI and 0.1\% PAL atmospheres for the UV and VIS channels. The broadband noise is indicated by the width of the lines. The magenta shaded regions represent phases where the exoplanet would be inside the inner working angle (IWA) of the LUVOIR A coronagraph, such that the coronagraph throughput significantly drops.}
    \label{Annual variability figure}
\end{figure*}

% Chemistry clouds albedo swap figure
\begin{figure}
	\centering
	\includegraphics[width=\columnwidth]{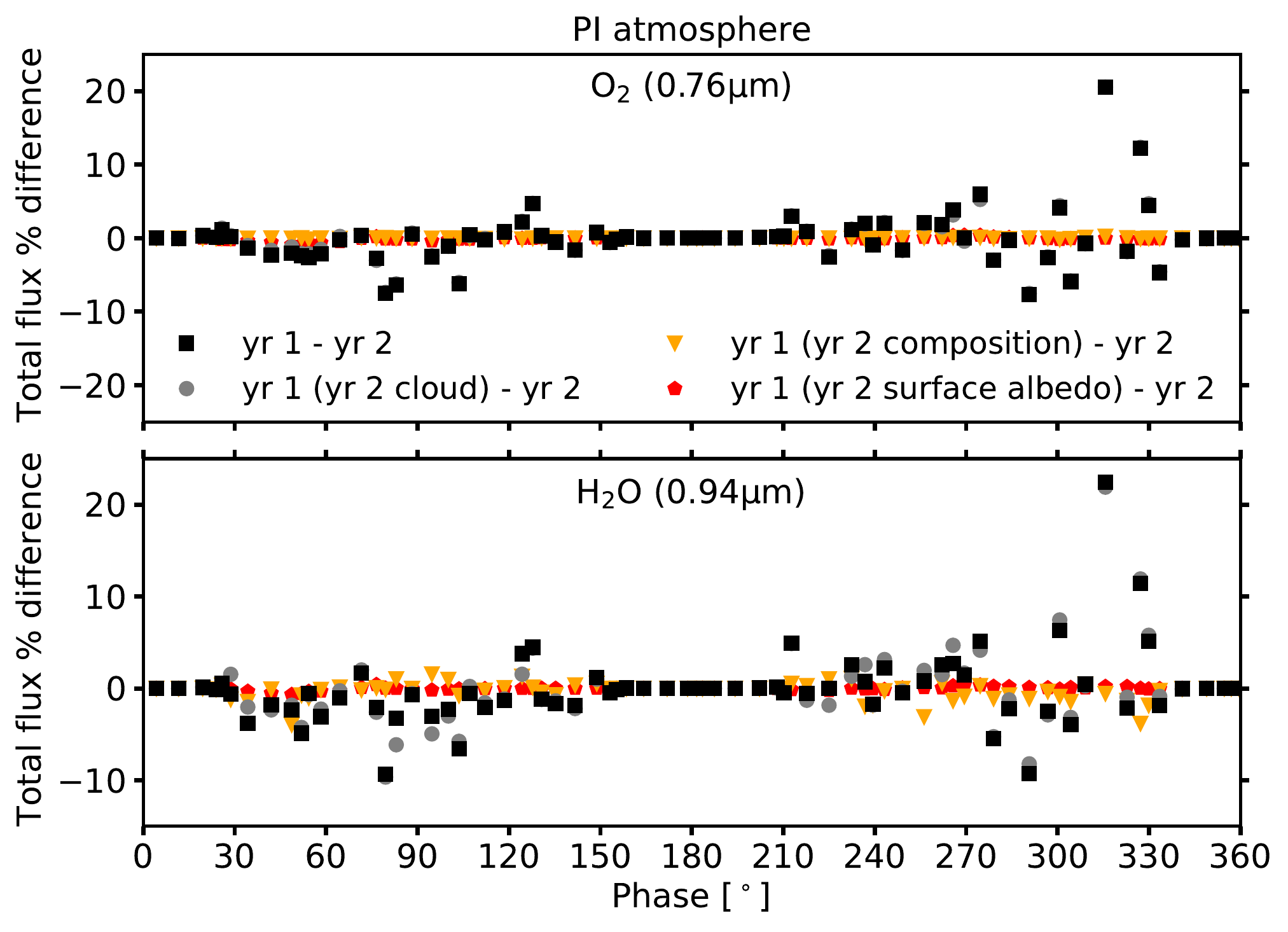}
    \caption{The total flux difference (in terms of percentage difference) versus orbital phase is plotted for the first and second year of WACCM6 data from the PI atmosphere for the \SI{0.76}{\micro \metre} \ce{O2} feature (\textbf{top}) and for the \SI{0.94}{\micro \metre} \ce{H2O} feature (\textbf{bottom}). Black squares indicate the difference between the standard output from the atmosphere with no changes. Grey circles, orange triangles, and red pentagons indicate the flux difference when clouds, composition, and surface albedo, respectively, from the second year, are swapped into the first year atmosphere before being processed by PSG. Although these swaps do not represent a physical atmospheric prediction, they enable quantification of what is affecting the atmospheric variability between years.}
    \label{Chemistry clouds albedo swap figure}
\end{figure}

% Range figure
\begin{figure*}
	\centering
	\includegraphics[width=1\textwidth]{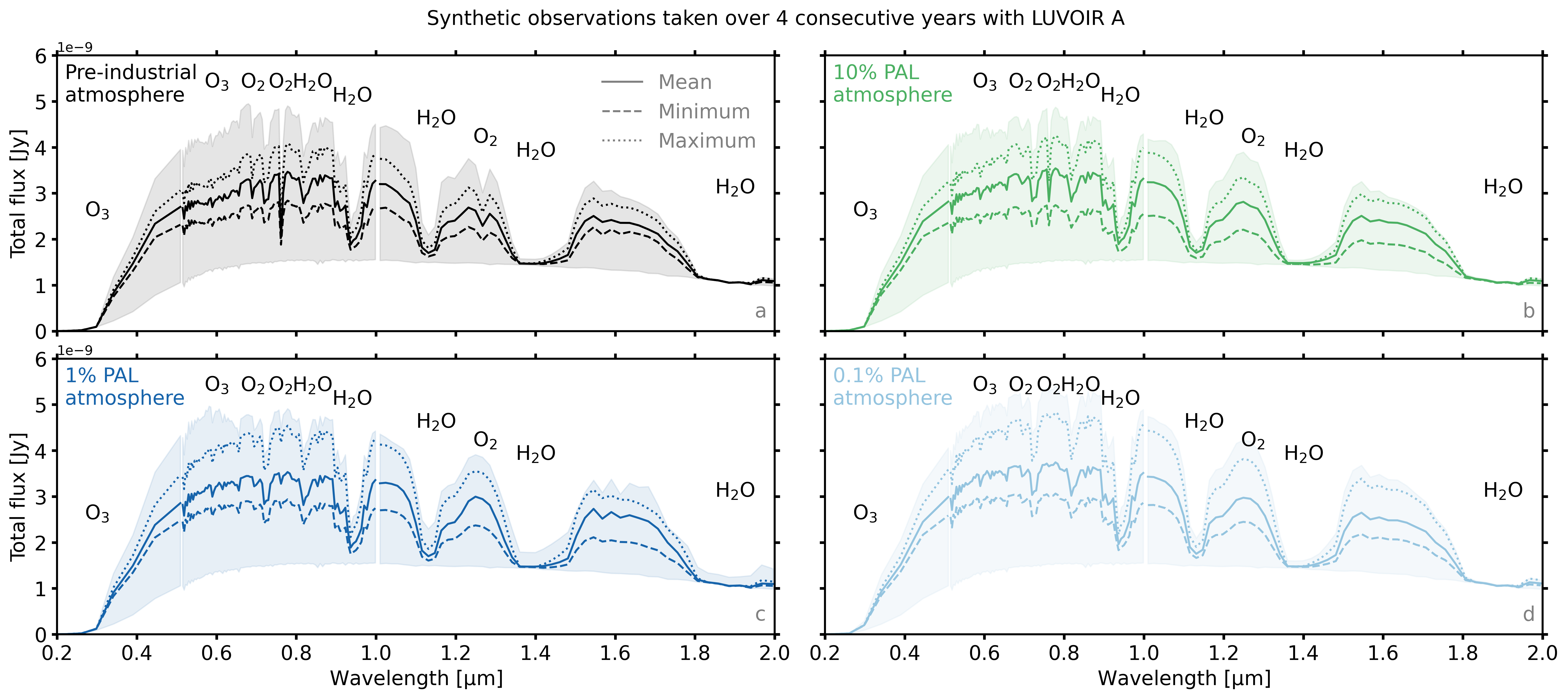}
    \caption{The mean (unbroken line), minimum (dashed line), and maximum (dotted line) for observations taken at maximum star-exoplanet separation (orbital phases closest to $90^\circ$ and $270^\circ$) are plotted for LUVOIR A at 10 pc. The minimum and maximum observations at quadrature are determined to be the date at which the \SI{0.88}{\micro \metre} line (often the wavelength with the highest flux) exhibits minimum and maximum flux for that particular year, respectively. The shaded regions show the range between the maximum and minimum total flux at each wavelength, now accounting for all phases. The results from the PI (\textbf{a}), 10\% PAL (\textbf{b}), 1\% PAL (\textbf{c}), and 0.1\% PAL (\textbf{d}) atmospheres are all plotted. Whilst maximum separation does not correspond to maximum planetary brightness, it does correspond to the most observable phase at different distances. Closer star-exoplanet separations may have reduced coronagraphic throughput.}
    \label{Range figure}
\end{figure*}

We present spectra generated using HCI for the PI and 0.1\% PAL atmosphere with the LUVOIR A telescope concept for an exoplanet at 10 pc, with 24 hr or 48 hr of integration time (annual variability curves for the 10\% PAL and 1\% PAL simulations are included in the Supplementary Information). We recognise that this is an optimistic distance at which an Earth-analogue exoplanet could be found, so we also include calculations at 25 pc and 50 pc. Previous work has used 10 pc for a variety of rocky exoplanet predictions \citep[e.g.][]{2011ApJ...738..184F,2021AJ....161..150C,2022arXiv220410041A} and so this distance also allows meaningful comparisons with those works. 

The PSG radiation output is in terms of planetary flux ($F_p$), stellar flux ($F_*$), and the total flux received by the detector ($F_T = F_p + F_*$). Note that $F_p$ cannot be empirically detected in such coronagraphic observations where some stellar light still reaches the detector. The error bars on $F_T$ depend on several sources of noise, as described in Section \ref{Methods section}. 

\subsubsection{Maximum signal-to-noise ratio}
\label{SNR section}

We quantify the maximum SNR possible with the three telescope concepts in Fig.~\ref{SNR bar chart figure}, at 10 pc, 25 pc, and 50 pc, after a total integration time of 24 hr. Occurrence rate estimates for Earth-analogue planets around G-type stars (usually denoted as $\eta_{\oplus}$) in the literature have various definitions \citep{2011ApJ...738..151C, 2012ApJ...745...20T,2013ApJ...766...81F, 2019MNRAS.487..246Z, 2021AJ....161...36B}, and the calculated rates typically vary between 0.01-2 per star \citep{2013PNAS..11019273P,2012ApJ...745...20T,2015ApJ...809....8B,2021AJ....161...36B}, with 2 being the optimistic end of estimates. According to the SIMBAD online database\footnote{\href{http://simbad.u-strasbg.fr/simbad/}{http://simbad.u-strasbg.fr/simbad/}}, there are 13 G0V-G9V stars closer than 10 parsecs and 2,830 G0V-G9V stars closer than 75 pc. This implies that an Earth-analogue exoplanet with a random orbital inclination should be found somewhere between 10 pc and 75 pc. This is a large range and should be kept in mind  when considering the results from Fig.~\ref{SNR bar chart figure}.

Assuming the threshold for detection of a molecule is a SNR of $\geq5$, then \ce{O3} in the UV can be detected for all atmospheres and telescopes for an exoplanet at 10 pc. \ce{O2} can be detected in the PI atmosphere by LUVOIR A at both \SI{0.69}{\micro \metre} and \SI{0.76}{\micro \metre}, although for the 0.1\% PAL atmosphere this would require an integration time of $\approx 50$ days and $\approx 1.2$ days for the \SI{0.69}{\micro \metre} and \SI{0.76}{\micro \metre} features, respectively. \ce{H2O} can be detected at \SI{0.94}{\micro \metre} with LUVOIR A but not at \SI{1.4}{\micro \metre}, and the LUVOIR B and HabEx SS cannot detect it at 10 pc with a 24 hr integration time. Rebinning of the two assessed \ce{H2O} features would improve their signal-to-noise ratios. At a distance of 25 pc, only LUVOIR A can detect any molecules (\ce{O3}). No molecules can be detected by any telescope in any atmosphere for an exoplanet at 50 pc, but \ce{O3} in the PI and 10\% PAL atmospheres may be detectable for $T_\textrm{int} \approx 3.4$ days with LUVOIR A. LUVOIR A (15 m) is the best performing telescope overall, and HabEx with a starshade (4 m) performs better than LUVOIR B (6 m) despite the larger diameter telescope size of LUVOIR B. For example, the SNR for \ce{O3} at 10 pc is up to a factor of 2.2 higher for HabEx with a starshade when compared to LUVOIR B.

Similar to \cite{2021AJ....161..150C}, where a cloud-free atmosphere was assumed in their calculations, we also find that the detection of \ce{O2} and \ce{O3} becomes increasingly difficult as the \ce{O2} concentration reduces. Because a different \ce{O2} concentration affects atmospheric chemistry and photolysis rates, the energy budget for the atmosphere changes. This results in different tropospheric temperatures which are positively correlated with the total water column (see the Supplementary Information for how the water column varies with orbital phase) and thus influences the reflection and emission spectra by modulating absorption and emission. Therefore, we find that the SNR for detecting \ce{H2O} is also influenced by the \ce{O2} concentration, although the change between atmospheres is small. The effect will be more significant for transmission spectra where the middle atmospheric concentration of \ce{H2O} reduces with reduced \ce{O2}.

Additionally, decreasing the diameter of cloud particle sizes from \SI{5}{\micro \metre} and \SI{100}{\micro \metre} to \SI{1}{\micro \metre} and \SI{1}{\micro \metre} (for water and ice clouds respectively) increases the maximum signal-to-noise ratio available during an orbit due to an increased amount of photons reaching the telescope detector. To give an example, the maximum SNR at 24 hr integration time with LUVOIR A for \ce{O3} at \SI{0.3}{\micro \metre} in the PI atmosphere is 66 with the larger diameter particles, versus a SNR of 100 with the smaller diameter cloud particles.

\subsubsection{Annual variations with LUVOIR A at 10 pc}

In Fig.~\ref{Annual variability figure}, we show how the same atmospheres exhibit inter-annual variability by modelling the spectra from HCI of four consecutive model years using the LUVOIR A telescope concept. This is the most optimistic future telescope concept we have assessed as discussed in Section \ref{SNR section}. We choose the final four years for each simulation as these data are how we have assessed the stability of the climate (see Section~\ref{The climate model WACCM6 section}). Fig.~\ref{Annual variability figure} shows the measured total flux ($F_T$) as a function of orbital phase, where January $1^\textrm{st}$ is at a phase of $279^\circ$. 

For LUVOIR A observing an oxygenated Earth analogue exoplanet at 10 pc, the inter-annual variability for the broadband and individual spectral features year-on-year is larger than the noise in the VIS channel at multiple phases. Specific lines within the UV do not show observable year-to-year variations, although the broadband in this channel does. Because the inner working angle of the coronagraph is proportional to wavelength, longer wavelengths are increasingly harder to observe due to reduced coronagraphic throughput - see Appendix \ref{Appendix coronagraph section}. This means that longer wavelengths, especially in the NIR channel, cannot be differentiated year-on-year. Additionally, certain phases cannot be differentiated, and this is either because the magnitude of the year-on-year variations are within the noise, or the coronagraph throughput is too low to detect any meaningful planetary flux variation. 

The \ce{H2O} and \ce{O2} features have larger variations compared to \ce{O3} features. Variations for \ce{H2O} and \ce{O2} at some phases are distinguishable from the noise for some years, but not necessarily all of the years. Therefore, multiple years ($\geq 3$) of observations may be needed to witness variations at particular phases. Greater integration times will increase the amount of phases where inter-annual variation is observable. Almost all phases can be differentiated in broadband observations.

In the UV channel, the same wavelength at the same phase between years can vary in brightness by up to 1.3 times and 2.4 times for $F_T$ and $F_p$, respectively. In the VIS and NIR channels, these numbers can reach up to 1.5 and 2.5, and 1.7 and 5, respectively.

% Phase shift figure LUVOIR
\begin{figure*}
	\centering
	\includegraphics[width=1\textwidth]{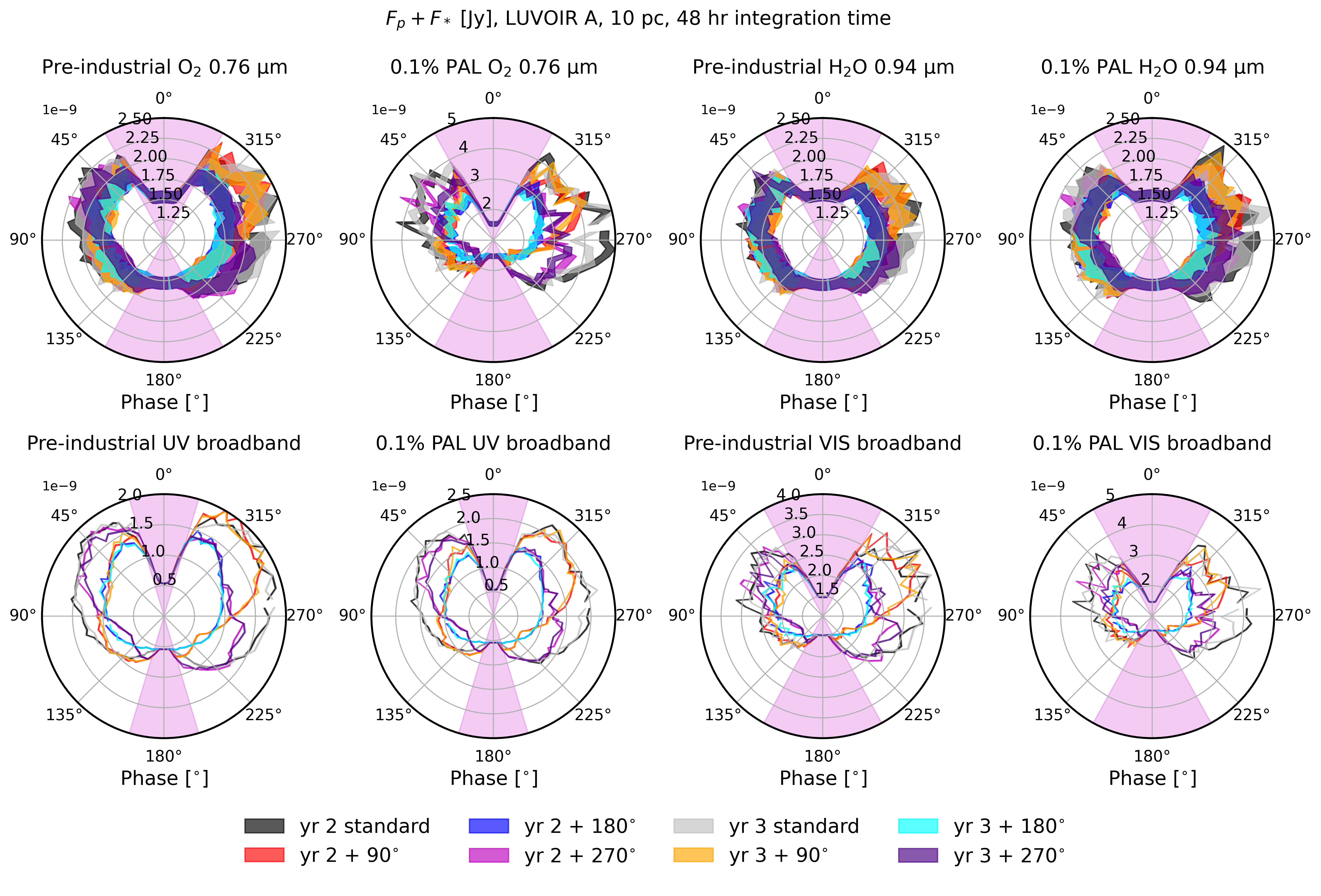}
    \caption{This figure demonstrates how seasons can impact exoplanet spectra. In units of Jy the total flux received by the telescope ($F_T = F_p + F_*$) is plotted in the radial direction against phase in the azimuthal direction. This is for a 48 hr integration time using LUVOIR A for an exoplanet at 10 pc. A standard year is plotted, and then observer geometry is rotated $+90^\circ, +180^\circ$ and $+270^\circ$. `yr 2' and `yr 3' refer to the $2^\textrm{nd}$ and $3^\textrm{rd}$ years of the final 4-year dataset, respectively. \textbf{Top}: Variation for the pre-industrial and 0.1\% PAL atmospheres are shown for \ce{O2} at \SI{0.76}{\micro \metre} and  \ce{H2O} at \SI{0.94}{\micro \metre}. The width of the lines represent the $\pm1$-$\sigma_\textrm{noise}$ uncertainty on the observations. \textbf{Bottom}: The broadband variation is shown in the UV and VIS channels with phase. The discontinuity in the curves is between December $27^\textrm{th}$ and January $1^\textrm{st}$. The magenta shaded regions represent phases where the exoplanet would be inside the inner working angle (IWA) of the LUVOIR A coronagraph. The width of the lines represent the $\pm1$-$\sigma_\textrm{noise}$ broadband uncertainty.}
    \label{Phase shift curve figure}
\end{figure*}

Fig.~\ref{Chemistry clouds albedo swap figure} shows the year 1 - year 2 percentage differences in the total flux versus orbital phase for the \SI{0.76}{\micro \metre} \ce{O2} and \SI{0.94}{\micro \metre} \ce{H2O} features. In addition, we swapped either clouds, composition, or surface albedo between the years to isolate the contribution of these planetary properties to the total year-to-year difference. The largest percentage difference for \ce{H2O} between these two years is +22.4\%, occurring at an orbital phase of $316^\circ$, where the percentage difference from only swapping clouds is +21.9\%. On the other hand, a swap in chemistry never contributes a change of more than $\pm4$\%, and surface albedo never more than $\pm1$\%. We can summarise this by stating that in our simulations, the inter-annual variation is primarily influenced by clouds, secondarily by chemistry (composition), and surface albedo plays only a minor role. This trend is similar across oxygenation states and wavelengths.

Total cloud coverage and cloud water path changes often for each atmosphere, such that clouds change the overall albedo of the planet. The \ce{O2} column does not significantly vary throughout an orbit, whereas the \ce{H2O} column does due to atmospheric temperature fluctuations and resultant changes in water phase. This is why a change in composition influences the spectra more for \ce{H2O} features than for \ce{O2} features. For a specific date and corresponding phase, year-to-year, the surface albedo is almost constant. In the cases considered here, it has little effect on observed annual variations. The surface albedo does affect the seasonal variations however, as we will show in the next section.

% \ce{H2O} features are more variable than the continuum, whereas \ce{O2} features are not.

In Fig.~\ref{Range figure}, we show how each atmosphere varies over 4 consecutive years at quadrature (maximum planet-star separation, where the orbital phase is either $90^\circ$ or $270^\circ$). The mean is plotted alongside the range. The largest positive percentage difference in flux from the mean flux at quadrature for each atmosphere is +18\%, +21\%, +29\%, and +30\%, for the PI, 10\% PAL, 1\% PAL, and 0.1\% PAL atmosphere, respectively. Likewise, The largest negative percentage difference is -18\%, -23\%, -23\%, and -19\%, for the PI, 10\% PAL, 1\% PAL, and 0.1\% PAL atmosphere, respectively.

We also show the total range between all phases and wavelengths (shaded background in Fig.~\ref{Range figure}). Although the signal-to-noise ratio for \ce{O3} is high in the UV, \ce{O3} column variations due to climate may not be detectable because of the very small range at \SI{0.3}{\micro \metre}. The total flux range varies with wavelength, and the largest total flux range occurs in the 0.1\% PAL case. The flux range of wavelengths which cover \ce{O3} and \ce{O2} features, including the continuum, generally increases with decreasing atmospheric \ce{O2} concentration. This trend with \ce{O2} concentration is not always linear for \ce{H2O} features. These results indicate that inter-annual variations in observations may be dependent on \ce{O2} concentrations, but more than four orbits would be required to determine this as a statistically reliable result, rather than a consequence of chemistry-independent climate variability within the model.

\subsubsection{Seasonal variations with LUVOIR A at 10 pc}
\label{Seasonal variations section}

Aside from inter-annual variations, observer geometry also affects the spectra as we will demonstrate in this section. We have varied the observer geometry simply by adding $90^\circ$, $180^\circ$, and $270^\circ$ to the orbital phase to which each model date corresponds to. We do this for the synthetic spectra for two consecutive model years (year 2 and year 3 from the final 4 years of data) to demonstrate how seasonal variations affect exoplanet spectra in Fig.~\ref{Phase shift curve figure}. 

We find that seasonal variations are larger than inter-annual variations between the same phase, as would be expected for a planet like Earth where the seasons are set by Earth's obliquity, i.e. the climatic differences between July $1^\textrm{st}$ and January $1^\textrm{st}$ are more likely to exhibit a greater variation in magnitude than two consecutive years on January $1^\textrm{st}$. Because one would likely not know a priori which season is being observed, this can initially limit the observational information gained about a particular exoplanet's climate because of the potential for degenerate interpretations. 

The reason why the phases between $180^\circ$-$360^\circ$ in Fig.~\ref{Annual variability figure} are often (but not always) brighter than $0^\circ$-$180^\circ$ by up to $\approx 15$\% is due to an increase in surface ice coverage. The \ce{H2O} column shows a seasonal cycle and this will affect the magnitude of the signal for \ce{H2O} features (e.g. more \ce{H2O} results in increased absorption). The influence from the cloud column is less obvious when comparing these two time periods because clouds do not vary with a seasonal cycle as clear as the ice seasonal cycle. $180^\circ$-$360^\circ$ roughly corresponds to September - March, which covers winter in the Northern hemisphere, when Arctic ice is most prevalent. The period between $0^\circ$-$180^\circ$ roughly corresponds to March - September, with increased surface ice around Antarctica during this time, although a higher proportion of the Earth's surface in the Arctic is covered in ice during the northern hemisphere's winter. More ice raises the albedo of Earth, such that the phases between $180^\circ$-$360^\circ$ are generally brighter. 

By shifting the phases that the dates correspond to, the same wavelength at the same phase for shifted seasons can vary in brightness by up to 2.8 and 10.5 times for $F_T$ and $F_p$, respectively, in the UV channel. In the VIS and NIR channels, these numbers can reach up to approximately 2.3 and 17, and 2 and 20, respectively.

\section{Discussion}
\label{Discussion}

We have demonstrated that chemistry and clouds impact the variability of high contrast imaging spectra of oxygenated terrestrial exoplanets. The variability depends upon the exoplanet's climate and chemistry, but the ability to detect this variation will be influenced by the distance, orbital geometry, and telescope design. Naturally, this will dictate what scientific results are retrievable from specific targets.

\subsection{Molecular detectability}

We acknowledge that the SNR estimates in section \ref{SNR section} depend on various parameters that we have assumed, such as a $90^\circ$ inclination angle, as well as the underpinning simulations from the Earth System Model WACCM6. Other chemistry-climate models when used in conjunction with PSG as we have here would yield different results \citep{2021arXiv210911460F}. The estimates also depend on telescope concepts that have not been constructed. Therefore, the results are an indicative, not exhaustive, picture of possible future measurements based on the information available for each concept at the time of writing. 

In the event that an Earth-like exoplanet is found even farther away (e.g. 75 pc) it would be much more challenging to detect any particular molecular feature. A longer integration time may not improve the performance because the coronagraph throughput significantly drops with distance for these concepts and the exoplanet would be observable only over a narrow range of phases and wavelengths, if any. The coronagraph throughput drops when the exoplanet passes inside the inner working angle of the coronagraph and is therefore partly or fully obscured by the coronagraph. The inner working angle for the LUVOIR concept is proportional to wavelength  so longer wavelengths are cut off and cannot be detected \citep{2019arXiv191206219T} - see Appendix \ref{Appendix coronagraph section}. 

The SNR of each feature is set in part by its depth relative to the continuum. It also depends on the telescope's diameter, the telescope's coronagraph, the total integration time, and distance to the planetary system. For example, even though the \ce{H2O} feature at \SI{1.4}{\micro \metre} is deeper than that at \SI{0.94}{\micro \metre}, the SNR is lower because of reduced coronagraphic throughput, and higher noise, in the near infrared. The HabEx with a starshade concept, when compared to a LUVOIR concept with an identically sized primary mirror, enables characterisation at greater distance due to the enhanced coronagraphic throughput  \citep{2021AJ....161..150C}.

\subsection{Predicted spectra variations from a 3D chemistry-climate model}

The \ce{O2} concentrations that we calculate in the lower and middle atmosphere are identical to those predicted by previous work \citep{1980JGR....85.3255K,2003AsBio...3..689S,2021E&PSL.56116818G}. However, we find that the lower and middle atmospheric concentrations of \ce{H2O}, \ce{O3}, \ce{CH4}, \ce{OH}, and other molecules, differ considerably compared to prior 1D and 3D modelling \citep{2022RSOS....911165C}. In this work we have considered the temporal variability of chemistry and climate and how this affects predicted emission and reflection spectra.

Although the effect of chemistry on the spectra for different oxygenation states is evident in Fig.~\ref{Ideal emission spectra figure}, chemistry has a smaller impact than when considering the time variability from the same simulation. The largest variation between years is caused by clouds, then chemistry, and then surface albedo. Nevertheless, surface albedo is important for seasonal cycles and is crucial for understanding the spectra presented in Section \ref{Seasonal variations section}. 1D atmospheric models which run to steady state cannot access this temporal variability, nor can they assess the seasonal effect on exoplanet spectra. Because clouds in 3D global climate models show significant variability between models  \citep{ceppi2017cloud,2013JCli...26.3823L,2020GMD....13..707F}, future comparisons with 3D time-resolved simulations using a different global climate model will therefore be of interest to determine how the predicted magnitude of spectral variations from HCI differ between models and oxygenation states (or various other chemical regimes). A comparison has been done for transmission spectra from 3D models in the TRAPPIST Habitable Atmosphere Intercomparison \citep{2021arXiv210911460F}. These comparisons could also be done with 1D models to evaluate the uncertainty across the entire range of models used in the exoplanet community, which would then be used to assign confidence to the interpretation of future observations.

Our simulated spectra were evaluated at 00:00 UTC, such that the Pacific ocean is illuminated, with a small proportion of the Earth's land and ice illuminated. The albedos of land and ice are both greater than that of the ocean, so reflection spectra from HCI could be affected if observations were calculated at different times. Additionally, the model time step is 30 minutes in our simulations, but we assumed this instantaneous state at 00:00 UTC held for long enough to integrate 24 hr or 48 hr of observations over. Our reasoning for this assumption is that saving multiple 3D variables every timestep for four model years and for multiple oxygenation states, then simulating each timestep snapshot with PSG (and with three telescope concepts), would require vastly more data storage and computational expense. As this assumption likely does not hold true \cite[see the diurnal variations predicted in low precision photometry measurements by][]{2001Natur.412..885F}, greater temporal variations may be introduced by further surface albedo and cloud variations. These variations during the timescale of the observations are likely to be important, especially when the observer does not know the rotation rate of the exoplanet. For instance, \cite{2020ApJ...893..140G} predicted how different planetary rotation rates affected the absorption depth of \ce{H2O}, \ce{O2}, and \ce{O3} features. Fortunately, the rotation rate may be known provided that approximately 5–15 rotations are observed, and the integration time is less than 1/6 to 1/4 of the planetary rotation period \citep{2022AJ....163...27L}.

The exoplanet is usually at its brightest just before secondary eclipse, but with the coronagraphs simulated here, such orbital phases are not observable as the planet is also being blocked by the coronagraph. Therefore, there is a trade-off between detectability and brightness. Other orbital phases show different ranges between minimum and maximum flux (see Fig.~\ref{Range figure}), so the observer may not be able to constrain certain parameters, including the amount of year-to-year variation across multiple phases, depending on the telescope used and the distance to the exoplanetary system.

With the mission concepts evaluated in this work, monitoring a nearby ($<25$ pc) Earth-analogue exoplanet for a single orbit is long enough to inform the observer about short-term climate variability and is ample time to detect if molecules of interest (e.g. \ce{O2}, \ce{O3}, and \ce{H2O}) are present. If inter-annual climate variability is of interest to future observers, then our simulations suggest it is observable provided conditions are favourable. Fully characterising an exoplanet (determining temperature, chemistry, seasons, obliquity, cloud coverage, rotation rate, land and ocean coverage, etc.) will be aided by observations of multiple orbits. Nevertheless, there will remain significant challenges to interpreting those observations.

\subsection{Challenges and future work}

Our atmospheric scenarios are based on the current understanding of Earth's atmospheric history since the GOE, when Earth's atmospheric oxygen concentrations rose above $10^{-5}$ PAL. However, the abundance of gases throughout Earth's history is not well constrained. Even if better constraints are achieved in the future, there is no reason why the atmospheric evolution of exoplanets that are oxygenated must take a similar path to Earth \citep{2021AGUA....200294K}. Nonetheless, the Earth system provides the best known template for an oxygenated atmosphere of a terrestrial planet. Large changes in the magnitude of the mixing ratios of other chemical species (e.g. \ce{CO2, CH4}), decreased or increased atmospheric pressure, and the proportion of land versus ocean coverage \citep{2022MNRAS.513.2761M}, will all affect our predictions. 

Changes in assumed cloud properties will also change the calculations presented in this work. \cite{2020ApJ...888L..20K} varied ice cloud particle radius between \SI{20}{\micro \metre} and \SI{200}{\micro \metre} and showed that this non-linearly affected the amount of transits required to detect water vapour in transmission spectra of a terrestrial exoplanet atmosphere using JWST. Although we have not covered a large range of cloud parameters, our results show that assumed particle sizes are important in PSG, affecting both variability and signal-to-noise ratios. Thus, the effects of how clouds are modelled in global climate models and radiative transfer models, and their impacts on predictions, requires further investigation.

There are many unknowns for exoplanets that compound the difficulty of modelling and interpreting spectra from a directly-imaged exoplanet. We have investigated a temporal and 3D problem that depends on: (1) albedo; (2) seasonal climatic variability \citep{2012ApJ...757...80C}; (3) clouds \citep{2018haex.bookE.116P}; (4) chemistry \citep{2018haex.bookE.116P}; (5) inter-annual climatic variability; (6) eccentricity \citep{2012ApJ...757...80C}; (7) obliquity \citep{2004NewA...10...67G, 2012ApJ...757...80C}; (8) observer inclination angle \citep{2012A&A...538A..95M, 2017A&A...601A.120B}; and (9) other exoplanets in the system contaminating the signal \citep{2018haex.bookE.116P}. Some of these variables interact with each other. Our high-contrast imaging spectra predictions account for some of these parameters (points 1-5) using a 3D fully coupled climate model for the first time, whilst others were not investigated in this work (6-9). Further work should investigate how these other parameters affect the predicted spectra of oxygenated exoplanets. At greater distances, where less orbital phases are available to observe, assessments of the year-to-year variability we demonstrate in our simulations will be less reliable. Additionally, our simulations assume that whilst $F_*$ varies with wavelength, the flux in each wavelength bin is assumed to be constant over the duration of an orbit. If we instead modified our simulations to include realistic stellar flux variations, it is unclear how detectable the observed flux variations due to the exoplanet would be.

Several methods already exist to obtain information that is otherwise difficult to access. For example, exo-cartography has been studied as a process for retrieving the surface maps of exoplanets, despite the planetary surface being unresolved in observations.  \citep{2018haex.bookE.147C,2019ApJ...882L...1F, 2019AJ....158..246B}. Already it is clear that different surfaces produce similar light curves, and lifting degeneracies to accurately map the surface of these exoplanets will be difficult. \cite{2022MNRAS.511..440T} used Earth observation data and found that for exoplanets with a similar climate to Earth, removing clouds to obtain a surface albedo map may produce unreliable results. They remarked that longer, continuous observing periods will help with mapping. Indeed, clouds may have the largest impact on spectra variation  \citep{2021arXiv210600079P}, as we have also found in our simualtions. Additionally, \cite{2016MNRAS.457..926S} found that, at least theoretically, it should be possible to determine the obliquity of an exoplanet by observing at 2–4 different orbital phases. With known obliquity, the impact of seasons can be estimated. Recently, \cite{2022MNRAS.tmpL..30P} used a machine learning algorithm - analysing more than 50,000 synthetic spectra of cold, Earth-like planets - to show that snow/clouds and water could be detected in 90\% and 70\% of cases, respectively. A similar evaluation using machine learning, and including both spectra produced from 1D and 3D atmospheric models, would be useful. Such a study would enable assessment of how reliably the community can use machine learning and 1D models to explore the parameter space that 3D models struggle to cover due to computational expense.

All of our simulations presume a biotic source of \ce{O2}. Even if a biogenic molecule is detected with high SNR, this may not be enough to prove that life exists on an exoplanet. Further information will likely be required for context to confirm that any potential biosignature that is detected has arisen from the presence of extraterrestrial life \citep{2017AsBio..17.1022M, 2018AsBio..18..709C, 2021AGUA....200294K}. This includes but is not limited to the detection of chemical disequilibrium \citep[e.g. \ce{O2} and \ce{CH4}][]{1965Natur.207..568L,1993Natur.365..715S,2016AsBio..16...39K}, or temporal changes in biogenic gases that indicate atmospheric modulation by life \citep{2018ApJ...858L..14O} rather than by geological processes. However, detecting such temporal chemical variations will likely be complicated by temporal changes in planetary albedo. We note here that detecting seasonal \ce{O3} variations in the UV as suggested by \cite{2018ApJ...858L..14O} may be complicated by clouds, as well as the seasons available to the observer, and because we have not explored all observing geometries, the effect of different inclination angles \citep{2018ApJ...858L..14O}.

Additionally, abiotic build-up of \ce{O2} is physically feasible on exoplanets \citep[and references therein]{2017AsBio..17.1022M}. Moreover, there is the possibility of detecting an exoplanet-exomoon system which has \ce{O2} and \ce{CH4} in the spectra, but not on the same celestial body \citep{2014PNAS..111.6871R}. In the coming decades, observational data, computational procedures which identify the sources of biologically-relevant gases \citep{2018AsBio..18..709C, 2020ApJ...898L..17L}, and simulations that utilise state-of-the art climate models, will all be needed to determine the most habitable exoplanets \citep{2019ApJS..243...30S}, and indeed to determine if any are inhabited.

Finally, we have only evaluated telescope concepts which will have spectroscopic capabilities between the ultraviolet and near infrared wavelength regions. To detect molecules such as \ce{O3}, \ce{H2O}, and \ce{CO2} in the infrared, this will require a mission design different to LUVOIR or HabEx. Proposed candidates include the Large Interferometer For Exoplanets \citep[LIFE;][]{2018SPIE10701E..1IQ, 2018ExA....46..543D} and the Origins Space telescope \citep[OST;][]{2021ExA....51..595W} missions. These telescope concepts may be able to constrain planetary emission in the mid-infrared and detect molecules such as \ce{H2O}, \ce{O3}, and \ce{CH4} \citep{2022arXiv220410041A}, potentially enabling the confirmation of chemical disequilibrium. Certainly, multi-wavelength observations which cover ultraviolet wavelengths to mid infrared wavelengths will enable a greater degree of atmospheric characterisation. Evaluating the OST and LIFE missions with our simulated atmospheres and calculating any variability that may be detected is left as future work.

\section{Conclusions}
\label{Conclusions section}

Using a state-of-the-art Earth System Model, WACCM6, we have simulated Earth-analogue exoplanets around a G2V star with varying concentrations of \ce{O2} (0.1\% PAL $\to$ 150\% PAL). Based on the Earth System Model output for these hypothetical exoplanets, we used the Planetary Spectrum Generator to simulate thousands of high-contrast imaging observations with three future telescope concepts: LUVOIR A, LUVOIR B, and HabEx with a starshade.

All telescopes are capable of detecting \ce{O3} at 10 pc after one day of integration time. \ce{H2O} and \ce{O2} detection may require longer integration times, or high \ce{O2} concentrations for improved SNR for \ce{O2} (e.g. $\geq10$\% PAL). At 25 pc, \ce{O3} in the UV can still be detected with LUVOIR A, but integration times longer 30 days may be required for \ce{H2O} and \ce{O2} detection.

Long term inter-annual climate variations and short-term variations can theoretically be observed in our simulated atmospheres with a telescope concept such as LUVOIR or HabEx. Both annual and seasonal variability can affect the brightness of key chemical features such as \ce{O2} and \ce{H2O}. The annual variability is primarily caused by clouds, and the variability appears to depend non-linearly on atmospheric \ce{O2} concentration (although more that four orbits would enable this to be a statistically robust result), as well as depending on wavelength, distance, and telescope design. The seasons that can be accessed through observations will affect the maximum magnitude of the incoming flux, with more ice coverage increasing this limit.

Future potential missions, such as the 6-metre diameter UV/VIS/IR telescope that was recommended by the Decadal Survey, and the LIFE mission, will present exciting opportunities to characterise the atmospheres of possible Earth-analogues. Confident interpretation of future observations that next-generation telescopes return will require sophisticated planetary modelling and observational analysis. It is important that future work continues to account for the complexity of chemistry and clouds, as well as long- and short-term climate variations.

\section*{Acknowledgements}

We would like to thank the anonymous reviewers for their helpful comments and suggestions.

G.J.C. acknowledges the studentship funded by the Science and Technology Facilities Council of the United Kingdom (STFC; grant number ST/T506230/1). C.W. acknowledges financial support from the University of Leeds and from the Science and Technology Facilities Council (grant numbers ST/T000287/1 and MR/T040726/1). This work was undertaken on ARC4, part of the High Performance Computing facilities at the University of Leeds, UK. 

We would like to acknowledge high-performance computing support from Cheyenne (doi:10.5065/D6RX99HX) provided by NCAR's Computational and Information Systems Laboratory, sponsored by the National Science Foundation. The CESM project is supported primarily by the National Science Foundation (NSF). This material is based upon work supported by the National Center for Atmospheric Research (NCAR), which is a major facility sponsored by the NSF under Cooperative Agreement 1852977.

\section*{Data availability}

The data underlying this article are available in the Dryad Digital Repository, at \href{https://doi.org/10.5061/dryad.cz8w9gj6f}{https://doi.org/10.5061/dryad.cz8w9gj6f}.

%%%%%%%%%%%%%%%%%%%%%%%%%%%%%%%%%%%%%%%%%%%%%%%%%%

%%%%%%%%%%%%%%%%%%%% REFERENCES %%%%%%%%%%%%%%%%%%

\bibliographystyle{mnras}
\bibliography{References}

% Don't change these lines
\bsp	% typesetting comment

\appendix

\section{Further details of completed PSG simulation}
\label{Appendix PSG simulations section}

We have not performed an exhaustive list of simulations in PSG. The parameter space is vast and to cover every future telescope concept, planet distance, oxygenation state, etc. is beyond the scope of this paper. See Table \ref{PSG run table} for the simulations that have been performed.

The PSG packages required for these calculations were \texttt{BASE}, \texttt{PROGRAMS}, and \texttt{CORRKLOW}. The versions used were last updated on 20/05/2022, 04/02/2022, and 01/03/2022, respectively.

Each simulation used $\textrm{NMAX}=3$ and $\textrm{LMAX}=41$. NMAX is defined as the number of stream pairs and the Legendre terms is given as LMAX - see the Fundamentals of the Planetary Spectrum Generator for more details \citep{2022fpsg.book.....V}. Calculation speed is proportional to LMAX but is approximately proportional to $\textrm{NMAX}^3$ \citep{2022fpsg.book.....V}. Multiple tests with different values of NMAX and LMAX were performed to ensure spectra contributions from scattering were accurate whilst not compromising the speed in such a way that collating the simulations would be unachievable within a reasonable time frame. In a small proportion of cases ($<2$\%) where unusually large brightness occurred, likely due to an asymmetry calculation in PSG, the substellar point was slightly shifted horizontally ($<3^\circ$) to produce consistent results.

\begin{table*}
\caption{For each of the PI, 10\% PAL, 1\% PAL, and 0.1\% PAL atmospheres, the PSG simulations that have been run are listed below. A single checkmark means that LUVOIR A has been evaluated in all three channels (UV, VIS, NIR) for 73 phase points around the orbit (one snapshot every 5 days), for a 24 hour integration time. Three checkmarks mean that simulated HCI observations for a full orbit for LUVOIR A, LUVOIR B, and HabEx with a starshade, have been evaluated. If they have also been evaluated at 25 pc and 50 pc, this is indicated in brackets. A dash means no telescope observations have been evaluated. A standard orbit is the where we have used PSG to calculate the total flux in Jy for each particular date and phase, using the ephemeris data from the year 2020. Signal-to-noise ratios (SNRs) have been calculated where a full orbit has been simulated in PSG for a particular atmosphere which has had all molecules removed apart from \ce{N2} and clouds. A phase shift orbit has a checkmark if spectra have been produced for the $+90^\circ$, $+180^\circ$, and $+270^\circ$ configurations. The water cloud particle size (WCPS) and the ice cloud particle size (ICPS) is indicated.}
\label{PSG run table}
\begin{tabular}{@{}ccccc@{}}
\toprule
 & year 1 & year 2  & year 3  & year 4  \\ \midrule

Standard orbit (\SI{5}{\micro \metre} WCPS, \SI{100}{\micro \metre} ICPS) & \checkmark \checkmark  \checkmark (10, 25, and 50 pc) & \checkmark (10 pc) & \checkmark (10 pc)  & \checkmark (10 pc) \\
Standard orbit SNR (\SI{5}{\micro \metre} WCPS, \SI{100}{\micro \metre} ICPS) & \checkmark \checkmark \checkmark  (10, 25, and 50 pc) & -  & -   & -  \\
Phase shift orbit (\SI{5}{\micro \metre} WCPS, \SI{100}{\micro \metre} ICPS)  & - & \checkmark (10 pc) & \checkmark (10 pc) & - \\

Standard orbit (\SI{1}{\micro \metre} WCPS, \SI{1}{\micro \metre} ICPS) & \checkmark \checkmark  \checkmark (10, 25, and 50 pc) & \checkmark (10 pc) & \checkmark (10 pc)  & \checkmark (10 pc) \\
Standard orbit SNR (\SI{1}{\micro \metre} WCPS, \SI{1}{\micro \metre} ICPS) & \checkmark \checkmark \checkmark  (10, 25, and 50 pc) & -  & -   & -  \\
Phase shift orbit (\SI{1}{\micro \metre} WCPS, \SI{1}{\micro \metre} ICPS)  & - & \checkmark (10 pc) & \checkmark (10 pc) & - \\

\end{tabular}
\end{table*}

\section{Observability of exoplanet using coronagraph}
\label{Appendix coronagraph section}

As an exoplanet orbits a star, with an orbital phase denoted by $\psi$, the coronagraph throughput varies because the projected exoplanet-star angular separation changes for the observer. When using a coronagraph, the Inner Working Angle (IWA) is typically defined as

\begin{equation}
    \textrm{IWA} = \frac{\lambda}{D},
\end{equation}

\noindent where $\lambda$ is the wavelength of light and $D$ is the diameter of the telescope. The telescope coronagraphs evaluated here are the LUVOIR ECLIPS instrument and HabEx with a starshade. For LUVOIR, the IWA is given in the LUVOIR Final Report \citep{2019arXiv191206219T} as $4\lambda/D$ for the UV channel, and $3.5\lambda/D$ for the VIS and NIR channels. For HabEx with a starshade, an IWA of 58 mas at 0.3–\SI{1.0}{\micro \metre} is given in the HabEx Final Report \citep{2020arXiv200106683G}. For the planet to be visible (and not blocked by the coronagraph), the IWA must be smaller than the angular separation of the star and planet ($\theta$), which is given by

\begin{equation}
    \theta \approx \frac{s}{d},
\end{equation}

\noindent where $s$ is the projected separation and $d$ is the distance to the planet-star system. The maximum projected separation for a circular orbit is the semi-major axis ($a$; 1 AU for Earth). For circular orbits of inclination, $i$, the separation, $s$, takes the general form 

\begin{equation}
    s = a\sqrt{\textrm{sin}^2(\psi)+\textrm{cos}^2(i) \textrm{cos}^2(\psi)}.
\end{equation}

\noindent In the simplified case of a system inclined at $90^\circ$ from the perspective of an observer,

\begin{equation}
    s = \pm a \textrm{sin}(\psi).
\end{equation}

\noindent When the IWA $<\theta$, the exoplanet is fully observable and when the IWA $>\theta$, the coronagraph throughput is reduced. The equations can be rearranged to find the orbital phase at which the exoplanet is about to enter inside the IWA, such that

\begin{equation}
    \psi =  \textrm{sin}^{-1} \left ( \frac{\lambda d}{a D} \right ) \textrm{for} \frac{\lambda d}{a D} \leq 1.
\end{equation}

\noindent Thus, one can see that as the wavelength $\lambda$ increases or the distance $d$ increases, then a smaller proportion of the orbit will be observable with the coronagraph. On the other hand, if the semi-major axis or the telescope diameter increases, then a larger proportion of the total orbit will be accessible for higher SNR observations.

\label{lastpage}

\end{document}